\documentclass[a4paper,12pt]{article}
\pdfoutput=1
\usepackage{jcappub} 
\usepackage{newtxtext,newtxmath}
\usepackage{color,xcolor}
\usepackage{lineno}
\usepackage{graphicx}	
\usepackage{latexsym,amsmath,amssymb}
\usepackage{multicol}
\usepackage{tikz}   
\usetikzlibrary{arrows}
\usepackage{subfigure,float}
\usepackage{dblfloatfix}
\usepackage{newtxmath}
\usepackage{subcaption}
\usepackage[export]{adjustbox}
\usetikzlibrary{shapes.geometric, arrows}
\graphicspath{{Figures/}} 
\usepackage{multicol}
\usepackage[normalem]{ulem}
\usepackage{parskip} 
\usepackage{soul,ulem}
\DeclareRobustCommand{\VAN}[3]{#2}
\let\VANthebibliography\thebibliography
\def\thebibliography{\DeclareRobustCommand{\VAN}[3]{##3}\VANthebibliography}
\usepackage{array}
\newcolumntype{P}[1]{>{\centering\arraybackslash}p{#1}}

\def \HI{{\sc Hi}}
\def \HII{{\sc Hii}}

\newcommand{\be}{\begin{equation}}
\newcommand{\e}{\end{equation}}
\newcommand{\bear}{\begin{eqnarray}}
\newcommand{\ear}{\end{eqnarray}}

\newcommand{\comment}[1]{}

\arxivnumber{2503.00919} 
\title{Interpreting the \HI\ $21$\,cm cosmology maps through Largest Cluster Statistics. Part II. Impact of the realistic foreground and instrumental noise on synthetic SKA1-Low observations}

\author[a,1]{Samit Kumar Pal,\note{Corresponding author.}}\emailAdd{palsamitkumar@gmail.com}
\author[b,c,a]{Saswata Dasgupta, }\emailAdd{saswata.iiti@gmail.com}
\author[a]{Abhirup Datta, }\emailAdd{datta.abhirup@gmail.com}
\author[a,d]{Suman Majumdar, }\emailAdd{mid.suman@gmail.com}
\author[e,f]{Satadru Bag, }\emailAdd{satadru.iucaa@gmail.com}
\author[g]{and Prakash Sarkar }\emailAdd{prakash.sarkar@gmail.com}
\affiliation[a]{Department of Astronomy, Astrophysics \& Space Engineering, Indian Institute of Technology Indore, Indore 453552, India}
\affiliation[b]{Institute of Astronomy, University of Cambridge, Cambridge, UK.}
\affiliation[c]{Kavli Institute for Cosmology,  University of Cambridge, Cambridge, UK.}
\affiliation[d]{Department of Physics, Blackett Laboratory, Imperial College, London SW7 2AZ, U. K.}
\affiliation[e]{Technical University of Munich, TUM School of Natural Sciences, Physics Department,  James-Franck-Stra{\ss}e 1, 85748 Garching, Germany}
\affiliation[f]{Max-Planck-Institut f{\"u}r Astrophysik, Karl-Schwarzschild Stra{\ss}e 1, 85748 Garching, Germany}
\affiliation[g]{Department of Physics, Kashi Sahu College, Seraikella, Jharkhand - 833219, India}

\date{\today}
\abstract{The Largest Cluster Statistics\,(LCS) analysis of the redshifted 21\,cm maps has been demonstrated to be an efficient and robust method for following the time evolution of the largest ionized regions\,(LIRs) during the Epoch of Reionization\,(EoR). The LCS can, in principle, constrain the reionization model and history by quantifying the morphology of neutral hydrogen\,(\HI) distribution during the different stages of the EoR. Specifically, the percolation transition of ionized regions, quantified and constrained via LCS, provides a crucial insight about the underlying reionization model. The previous LCS analysis of EoR 21\,cm maps demonstrates that the convolution of the synthesized beam of the radio interferometric arrays, e.g. SKA1-Low with the target signal, shifts the apparent percolation transition of ionized regions towards the lower redshifts. In this study, we present an optimal thresholding strategy to reduce this bias in the recovered percolation transition. We assess the robustness of LCS analysis of the 21\,cm maps, considering the effects of antenna-based gain calibration errors and instrumental noise for SKA1-Low. This analysis is performed using synthetic observations simulated by the \textsc{21cmE2E} pipeline, considering SKA1-Low AA4 configuration within a radius of 2\,km from the array centre. Our findings suggest that a minimum of $2000$\,hours of observation (SNR $\gtrapprox 3$) are required for the LCS analysis to credibly suppress the confusion introduced by thermal noise. Further, we also demonstrate that for a maximum antenna-based calibration error tolerance of $\sim 0.02\%$ (post calibration), the reionization history can be recovered in a robust and relatively unbiased manner using the LCS.}    

 
\begin{document}
 \maketitle
 \flushbottom
\section{Introduction}\label{sec:intro}
The Cosmic Dawn\,(CD) and Epoch of Reionization\,(EoR) mark a significant period in the cosmic timeline, yet it remains one of the least understood epochs in the evolution of the universe. During this epoch, the UV and X-ray photons started to ionize the neutral hydrogen in the intergalactic medium (IGM). Studies of the absorption spectra of Ly$\alpha$ from high-redshift quasars suggest that reionization was nearly complete by $z\sim 6$ \citep{Fan_2006}. Our present knowledge suggests that the process of reionization began when the first stars and galaxies formed in the over-density regions. This process led to the ionization of nearby gas and the formation of individual ionized bubbles. However, the exact details of the reionization process, such as the nature and properties of the ionizing sources and the morphology and topology of the ionized bubbles at different stages of reionization, still remain uncertain.

The redshifted \HI\ $21$\,cm signal arises from the spin-flip transition of electrons in the ground state. The observation of the redshifted \HI\ 21\,cm signal serves as a direct window into the state of hydrogen in the IGM and, thereby, can potentially be used to study this complex period. By measuring spatial fluctuations in the 21\,cm signal with radio interferometry, it is possible to create tomographic maps of \HI\ regions throughout the sky. The detection of the redshifted \HI\ 21\,cm signal from CD/EoR is a pivotal science goal of first-generation radio interferometers such as GMRT \citep{Paciga_2013}, MWA \citep{Kolopanis_2023}, LOFAR \citep{Mertens_2020}, and HERA \citep{Hera_2023}. Due to the low signal-to-noise (SNR) ratio, these interferometers were focused on statistical detection of the target signal, such as the power spectrum. Next-generation radio interferometers such as HERA and SKA-Low are expected to precisely measure the \HI\ 21\,cm signal power spectrum (PS) from CD/EoR with high precision. However, the detection of 21\,cm signal is very challenging because of the bright astrophysical foreground, which is $4-5$ orders of magnitude brighter than the target signal \citep{Bharadwaj2005, Jelic2008, Jelic2010, Zann2011, Choudhuri2017, Chakraborty2019, Mazumdar2020}. In addition, along with foreground, calibration errors\citep{Barry2016, Patil2016, EwallWice2017, Mazumder_2022}, ionospheric disturbances \citep{Jordan2017, Trott2018, Pal2025}, and instrumental effects \citep{Kern2019} introduce distortion in the target signal.

Although the PS is a powerful tool, the EoR 21\,cm signal is expected to be strongly non-Gaussian, and the PS alone could not fully describe it. Therefore, higher-order statistics, such as the bispectrum \citep{Majumder_2018, Kamran_2022} and trispectrum\citep{Cooray2008} are necessary to capture this non-Gaussian nature. The image-based e.g. statistics Minkowski functionals (MFs) \citep{Gleser_2006, Lee_2008, Friedrich_2011, Hong2014, Yoshiura_2017, Bag_2018, Bag_2019, Spina2021, Pathak_2022, Dasgupta2023} and Minkowski Tensors also provide a useful way to explore this morphological and topological evolution of reionization. Additionally, Percolation theory \citep{Ilieve_2006, Iliev_2014, Furlanetto_2016, Bag_2018, Bag_2019}, granulometry \cite{Kakiichi2017}, persistence theory \cite{Elbers2019}, and Betti numbers \cite{Giri_2021, Kapahtia2021} are some of the methods employed to analyze the topological phases of ionized hydrogen (\HII) regions during the EoR.

It is generally accepted that conclusions from these image-based methods depend on detecting a large number of \HII\ regions across different stages and sizes. However, the study by Bag et al. \citep{Bag_2018, Bag_2019} demonstrates that identifying only the largest ionized region\,(LIR) is sufficient to infer the percolation process. To reach this conclusion, they introduced a novel statistic named Largest Cluster Statistics\,(LCS), along with a shape-finding algorithm. The LCS analysis of the redshifted 21\,cm maps has been demonstrated to be an efficient and robust method for tracking the time evolution of the LIRs during the EoR. The LCS can constrain the reionization model and history by quantifying the morphology of neutral hydrogen distribution during the different stages of the EoR. Specifically, the percolation transition of ionized regions, quantified and constrained via LCS, provides a crucial insight about the underlying reionization model \citep{Pathak_2022}. Our previous work, Dasgupta et al. \citep{Dasgupta2023}, demonstrated how the convolution of the synthesized beam of the radio interferometric arrays, e.g. SKA1-Low with the target signal, can affect our conclusion about the percolation of \HII\ regions during reionization. They showed that the apparent percolation transition of ionized regions shifted towards the later stage of reionization, depending upon the array synthesized beam of SKA1-Low and thresholding formalism used in noisy data.

Motivated by this, we present an optimal thresholding strategy to minimize the bias in the recovered percolation transition. We further assess the robustness of our LCS analysis of the 21\,cm maps under various foreground contamination scenarios. For this purpose, we consider two sources of foreground contamination: a) extra-galactic point sources and b) diffuse synchrotron and free-free emission. For all of these sky models, we consider the antenna-based gain calibration errors to estimate the maximum tolerance level needed to recover the reionization history in a robust and relatively unbiased manner using LCS. In addition, we have also investigated the impact of instrumental noise for SKA1-Low on this analysis via the synthetic observations simulated by the \textsc{21cmE2E} pipeline\footnote{\url{https://gitlab.com/samit-pal/21cme2e}}. This analysis is done by diagnostic tool \textsc{SURFGEN2} \citep{Bag_2018, Bag_2019} to estimate the LCS and gain insights into the percolation of \HII\ regions during the EoR.

This paper is organized as follows: In Section~\ref{sec:sky_model}, we discuss the simulation of the radio sky. Section~\ref{sec:sim} describes our end-to-end simulation and the methodology used to incorporate antenna-based gain calibration errors and instrumental noise on synthetic SKA1-Low observations. The analysis formalism is presented in Section~\ref{sec:method}, followed by the results in Section~\ref{sec:results}. Finally, we summarize and discuss our findings in Section~\ref{sec:summary}. We used the best-fitted cosmological parameters from the WMAP five-year data release that have been used throughout the paper, which details as follows: $h = 0.7, \Omega_{m} = 0.27, \Omega_{\Lambda} = 0.73, \Omega_{b}h^2 = 0.0226$ \citep{Komatsu2009}. 

\section{Simulations of the radio sky}\label{sec:sky_model}
We investigate the robustness of LCS analysis to study the evolution of the largest ionized region during different stages of reionization. This analysis uses synthetic SKA1-Low observations generated from a realistic simulation based on \textsc{21cmE2E}-pipeline. This section describes the simulation of sky models. The sky models consist of the \HI\ signal and the astrophysical foreground within redshift range $7.2<z<8.8$, corresponding to frequencies $\sim 144 - 173$\,MHz.

\subsection{Seminumerical simulation of reionization}\label{sec:semi_sim}
In this section, we provide a brief review of the simulation of the \HI\ fields at different stages of the EoR. For detailed information, readers can refer to Section 2 of Dasgupta et al. \cite{Dasgupta2023}. To construct the brightness temperature maps of \HI\ 21\,cm signal, we used the {\sc ReionYuga} simulation \citep{Majumdar_2014, Majumdar_2016, Mondal_2017}. This simulation employs a semi-numerical approach based on the excursion set formalism. The {\sc ReionYuga} utilizes an N-body simulation to create the distribution of dark matter at a given redshift. Next, a Friends-of-Friends (FoF) halo finding algorithm was used to detect the occurrence of the collapse of dark-matter halos inside this distribution of matter. The first light sources, which emit reionizing photons, are formed halos. The ionization fields created via excursion set formalism are thereafter transformed into the field of $21$\,cm brightness temperature. For our analysis, we used the existing simulated \HI\ $21$\,cm maps from Dasgupta et al.\cite{Dasgupta2023}. These maps are coeval boxes, where each box measures $143.36$\,cMpc on each side and is distributed over a mesh consisting of a $256^3$ grid volume. A detailed study of the evolution of the LCS along a lightcone to understand how the lightcone effect biases the percolation curve and affects the distinguishability of source models is presented in Potluri et al. (in prep.).

\subsection{Astrophysical foregrounds}
\label{foregrounds}
One of the major contaminants of CD/EoR experiments is the astrophysical foreground. Its brightness is $4-5$ orders of magnitude higher than the faint $21$\,cm signal. The primary components in the foreground include diffuse galactic synchrotron radiation, galactic and extra-galactic free-free radiation and extra-galactic point sources. The foreground emission is expected to be spectrally smooth. However, calibration and instrumental effects introduce additional unsmooth structures. We considered two sources of foreground contamination, diffuse emission and extra-galactic point sources, to test these impacts on the residual maps. The foreground contributions are detailed below.

\subsubsection{Diffuse emission} 
\label{diffuse_emission}
The diffuse galactic emissions is dominated by the synchrotron and free-free emissions. Being large-scale structures, these are sensitive to shorter baselines. The diffuse synchrotron and free-free emission foregrounds were simulated using models outlined in \citep{Carucci_2020, Cunnington_2021}. The Planck Sky Model (PSM) at $217$\,GHz are used to simulate the diffuse maps. At each pixel $p$, the brightness of these emissions is quantified using brightness temperature $T_s$. These are modelled as power laws:
\begin{equation}
    T(\nu,p) = T_s \left(\frac{\nu}{\nu_0} \right)^{\beta_{s}(\nu,p)}
    \label{equ:power_law}
\end{equation}
where $\beta_{s}$ is the spectral index. The all-sky amplitudes for synchrotron and free-free emissions simulations are publicaly available on the Planck Legacy Archive\footnote{\url{https://pla.esac.esa.int/\#maps}}. Using the {\sc healpix} routines, we degrade and smooth these maps to our desired {\sc nside}\footnote{The resolution of the map is defined by the {\sc nside} parameter} and resolution. The amplitude of diffuse emission is determined from the PSM at $217$\,GHz with a resolution of {\sc nside}= 2048. To determine the spectral index map of synchrotron emission, we used $217$\,GHz and $353$\,GHz synchrotron maps at {\sc nside}=2048. Therefore, the spectral index varies in each pixel for synchrotron maps, whereas free-free emission $\beta_{\text{ff}} = -2.13$ \citep{Cunnington_2021}, is constant in all the pixels. These spectral indices and brightness temperature maps obtained at $217$\,GHz are used to extrapolate to the frequencies of interest using the equation \ref{equ:power_law}. Here, we simulated the diffused foregrounds at $\alpha = 0$\,h and $\delta = -30^{\circ}$ field and rotated the centre of the field to our pointing centre. 

We generate image cubes by including the calibration error to explore the robustness of our LCS analysis. A more detailed discussion is provided in Section~\ref{sec:di_gain_cal}. In a synthetic SKA1-Low observation, the residual contamination left in the image cube will impact the evolution history of reionization. Following Mazumder et al. \citep{Mazumder_2022}, we vary the foreground residual amplitude from $10^{-3}\%$ to $10^{-2}\%$ of the actual sky emission of our simulation. This variation is used to study its impact on the estimated LCS evolution.

\subsubsection{Extragalactic point sources}
\label{point_source}
In addition to the galactic and extra-galactic diffuse emission,  the total foreground emission is also affected by extra-galactic point sources. These extra-galactic point sources are often compact and finite in size. This study uses the Tiered Radio Extra-galactic Continuum Simulation (T-RECS)\citep{Bonaldi_2019} catalogue. At a frequency of $150$ MHz, the flux values of the sources range from $3.1$ mJy to $0.6$ Jy. The fluxes were transformed to extrapolate to the frequencies of interest using the relationship $S_{\nu} \propto \nu^{-\alpha}$, where $\alpha$ is $-0.8$. These extra-galactic point sources mainly comprises of star-forming galaxies and radio-quiet quasars. For a detailed description of the T-RECS catalogue model, readers can refer to \citep{Bonaldi_2019, Mazumder_2022}. The T-RECS catalogue comprises of 2522 unpolarized flat-spectrum sources within the $(4^{\circ})^2$ sky area. However, we generated image cubes only for the central region, covering $(\sim 1.5^\circ)^2$ fields. This was done to match the field of view (FoV) of the input of the input \HI\ maps from {\sc reionyuga}, depending on the redshift of observation. Bright sources in the beam sidelobes present challenges for data calibration in real observations. However, these effects were not considered in this work.

\begin{figure}
    \centering
    \includegraphics[width =\linewidth]{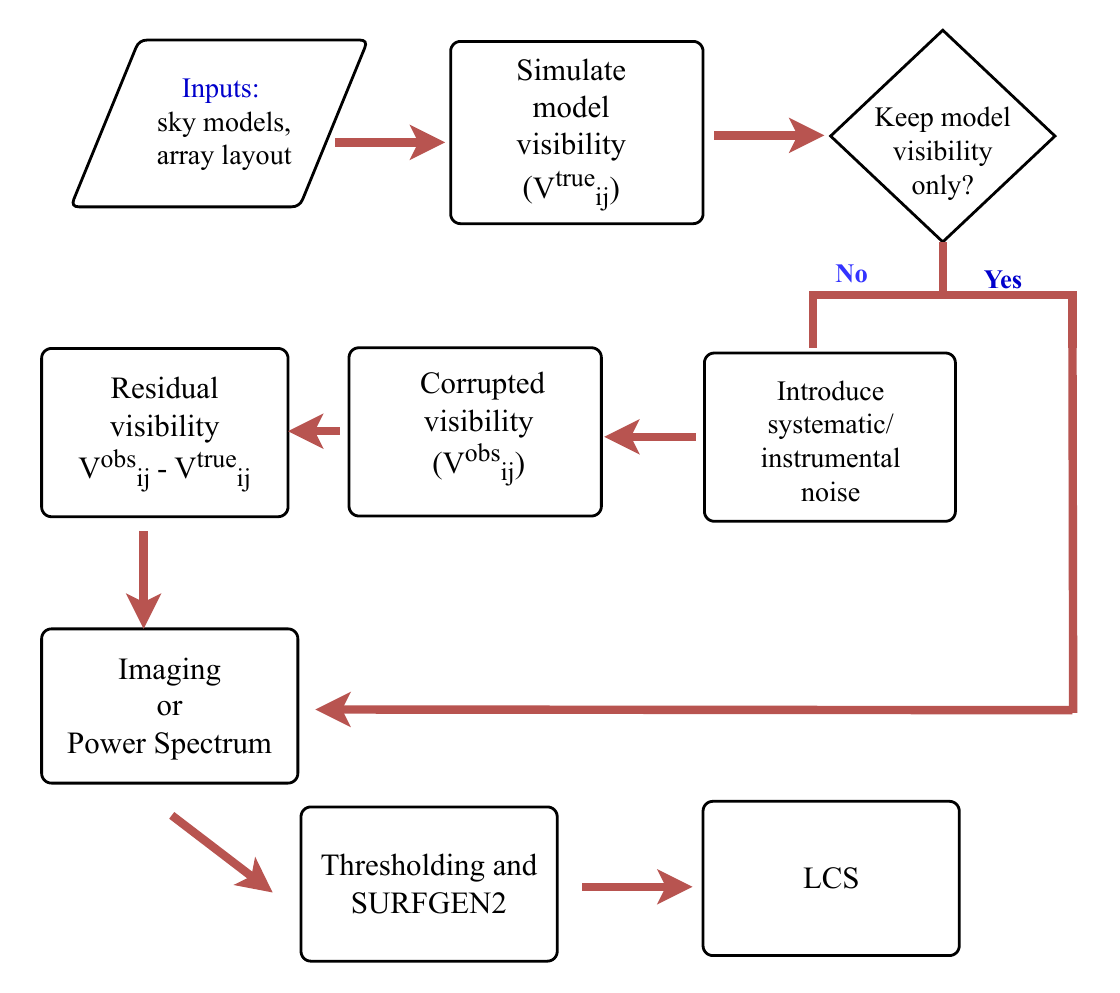}
    \caption{Schematic diagram of the \textsc{21cmE2E}-pipeline based on OSKAR and CASA software. This pipeline is used to estimate LCS from 21\,cm observation results.}
    \label{fig:e2e}
\end{figure}

\section{Simulations of observations}\label{sec:sim}
This section outlines the synthetic observation with SKA1-Low to simulate the radio sky discussed in Section \ref{sec:sky_model}. The observation parameters are listed in table~\ref{tab:obs_para}. Figure~\ref{fig:e2e} shows the schematic diagram of the \textsc{21cmE2E} pipeline. This pipeline is based on the \textsc{OSKAR}\footnote{\url{https://ska-telescope.gitlab.io/sim/oskar/}}. The \textsc{OSKAR} \citep{Dulwich2009} package generates the simulated visibility based on the input sky model, observational parameters, and telescope specifications. We simulated only $2$\,hours of observation, which will be restricted to  $\pm 1$\,hours around the transit time of the target EoR field for this study to reduce computational costs. The field of view centred at $\alpha = 15$\,h and $\delta = -30^{\circ}$. The integration time for the simulation was set to 10\,seconds. In the simulated \HI\ $21$\,cm cube, one axis represents the line-of-sight (LoS) or frequency axis. The cubes were then divided into slices based on their frequency resolution. Each slice was assigned a frequency label corresponding to the comoving distance from the observer. The slices of the $21$\,cm maps were converted from comoving Mpc (cMpc) to angular coordinates on the sky plane. These angular maps were then processed through the \textsc{21cmE2E} pipeline. We use \textsc{wsclean} to produce the image cube from simulated visibilities using the Briggs weighting scheme with robust parameter $0.4$. The frequency-labeled slices were stacked, and final image cubes were made for further analysis. It is noted that the observed \HI\ maps have the same size as the input \HI\ maps. The pixel size was chosen to match the size of the input \HI\ maps.

\begin{table}
    \centering
     \caption{An overview of the observational parameters used in the simulations.}
    \label{tab:obs_para}
    \begin{tabular}{lcccr}
    \hline\hline
    & \\
    Parameter &&  Value \\
           & & \\
    \hline
    Redshift range  && 7.2 - 8.8 \\
    Telescope && SKA1-Low AA4 \\
    Number of stations && 296 \\
    Maximum baseline       & &$\sim 3.15$\,km\\
    Polarization            & & Stokes I\\
    Phase Center(J2000)     && RA, DEC= 15\,h, $-30\,^{\circ}$\\
    Duration of obseravtion && 2\,h ($\pm 1$\,HA) \\
    Time resolution && 10\,s \\   
    Sky model && 2522 point sources in the central $4^{\circ}\times4^{\circ}$ \\
    \hline\hline
    \end{tabular}
 \end{table}

\subsection{Telescope model}\label{sec:tel}
\begin{figure}
    \centering
    \includegraphics[width=\linewidth]{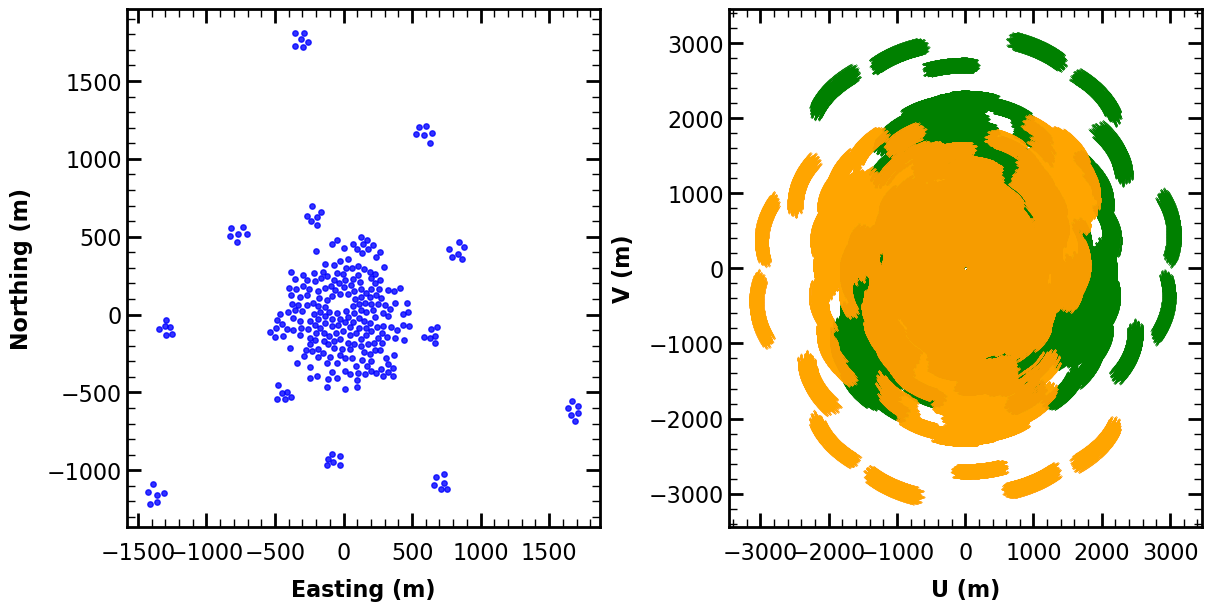}
    \caption{Left-hand panel: Telescope layout of the SKA1-Low array assembly 4\,(AA4) configuration within a radius of 2\,km from the array centre. Right-hand panel: Baseline coverage in the UV plane for an observation time of 2 hours ($\pm 1$\,HA). The U, V and -U, -V are plotted here using different colours for visual clarity. 
 }
    \label{fig:telescope}
\end{figure}

The SKA1-Low is one of the most sensitive upcoming radio interferometers. It is expected to make tomographic maps of the \HI\ 21\,cm signal from the EoR \cite{mellema2015}. The construction of the SKA1-Low radio interferometer is progressing rapidly in Inyarrimanha Ilgari Bundara,  Western Australia. The telescope will consist of $512$ stations with a maximum baseline of approximately $74$\,km. Since the EoR signal is mostly present on shorter baselines, it corresponds to large angular scales. We used the array assembly 4 (AA4) configuration of SKA1-Low within a radius of 2\,km from the array centre and exludes the longer baselines \citep{SKAO_telescope}. This compact array configuration consists of $296$\,stations. The station layout of SKA1-Low is shown in Figure \ref{fig:telescope}. 

\subsection{Gain calibration error}
\label{sec:di_gain_cal}
The main observable quantity of the radio interferometers is the visibility. However, non-ideal radio interferometers do not directly measure the true visibility, $V^{\rm true}_{\rm ij}$, for each baseline formed by the $i$th and $j$th antennas. Instead, they measure the observed visibility, $V^{\rm obs}_{\rm ij}$. The relationship between the observed and true visibility (assuming implicit time and frequency dependence) is given by
\begin{equation}
    V^{\rm obs}_{\rm ij} = g_{i}g^{*}_{j}V^{\rm true}_{\rm ij} + n_{\rm ij}
    \label{equ:calibration}
\end{equation}
Where $g_i$ \& $g^{*}_j$ is the complex antenna gain term and $n_{ij}$ is the noise on ${\rm ij}$th baseline. However, the true visibility, $V^{\rm true}_{\rm ij}$, can be measured by a perfectly calibrated noiseless interferometer. To obtain the true visibility, we need to calibrate the observation. During calibration, we solve the gain factors $g_{i}$ \& $g_{j}$ to minimize the chi-square between true and observed visibility. However, the accuracy of calibration is limited by the SNR. In an ideal scenario, the gain factor should be unity. However, an analytical solution of Equation~\ref{equ:calibration} for the desired parameter is intractable. Given $N$ antennas, there are $N(N-1)/2$ baselines. These baselines correspond to $N(N-1)/2$ true visibilities, which we want to solve along with $N$ antenna gain factors. This process of correcting the antenna gain is known as self-calibration or direction-independent calibration.  In $21$\,cm experiments, either sky-based or redundant calibration techniques are used. The accurate calibration is essential because the uncalibrated part of gains or time and frequency-correlated residual gains will propagate into the subsequent steps, thereby confusing or even obscuring during the extraction of the target signal. The complex gain from the antenna `i' can be modelled by
\begin{equation}
    g_{i} = (1+ \delta a_{i})e^{-i\delta \phi_i}
    \label{gain_factor_err}
\end{equation}
where $\delta a_{i}$ \& $\delta \phi_{i}$ are the errors in amplitude and phase. The $\delta a_{i}$ is dimensionless and $\delta \phi_{i}$ is measured in degree. The efficiency of the calibration is characterized by minimizing of $\delta a_{i}$ and $\delta \phi_{i}$. Due to the faint nature of $21$\,cm signal, achieving an $\text{SNR}\geq 1$  becomes a daunting challenge. This work quantifies the degree of accuracy needed for the LCS analysis for future SKA1-Low observations to study the percolation process during reionization.

To simulate antenna-based gain calibration errors, we added a noise term to the gain model with a specified standard deviation in amplitude and phase for both time and frequency domains (see equation \ref{gain_factor_err}). The calibration errors are sampled randomly with different standard deviations at the level in the range from $10^{-3}\%$ to  $10^{-2}\%$ in amplitude and from $10^{-3}$ to  $10^{-2}$\,degree in phase. The amplitude and phase response of the residual gain errors value for each station changes at every $10$\,second time stamp and every frequency channel throughout the observation. The mean amplitude response is set to unity and the mean phase response to zero, ensuring no systematic calibration offsets are introduced. To simulate realistic calibration errors, we assume that time and frequency fluctuations are independent, and the gain error at each time and frequency point is given by the complex product of two one-dimensional distributions- one for time and one for frequency. These one-dimensional gain distributions are initially populated with random samples from a Gaussian distribution, with specified standard deviations for both amplitude and phase. We then apply the \textsc{colorednoise}\footnote{\url{https://github.com/felixpatzelt/colorednoise.git}} Python code to transform these white-noise distributions, which have equal power at all scales, into red-noise distributions with a $-2$ power-law index. This process enhances the fluctuation power on the longest timescales and across the broadest frequency ranges while preserving the original mean and standard deviation. These final gain errors closely resemble the correlated error patterns observed in real calibration data. This approach to simulating direction-independent calibration errors was introduced in the SKA Science Data Challenge 3a (SDC3a) \cite{Bonaldi_2025arXiv}. The gain correlations are usually at time scales of a few tens of seconds \citep{Mangla2022, Mangla2023}. Since any such calibration procedure is done over short time scales and narrow frequency bands, we assume that residual calibration errors are not correlated beyond each night/day’s observations (i.e. 2\,hours). So, any systematic errors are only restricted within the 2\,hours of observing time. Hence, the RMS noise floor achieved after each epoch (2\,hours) of observations is then co-added with other epochs, and the RMS noise reduces as $1/\sqrt{N_{\rm days}}$, where $N_{\rm day}$ is the number of independent observing epochs. The assumptions made in our study are consistent with the previous work by Datta et al. \cite{Datta2009}. We focus on deep observations with a total duration of 1000 hours. Our simulation models this deep integration by assuming 500 times repetition of multiple nights of continuous 2-hour tracking observations of the same patch. The final post-calibration and post-averaging standard deviation values are then applied to each corresponding single observation period. This work does not apply any mitigation techniques, as the goal is to estimate the accuracy of any such techniques that will allow us to detect the EoR signal.

\begin{figure}[h!]
    \centering
    \includegraphics[width=\linewidth]{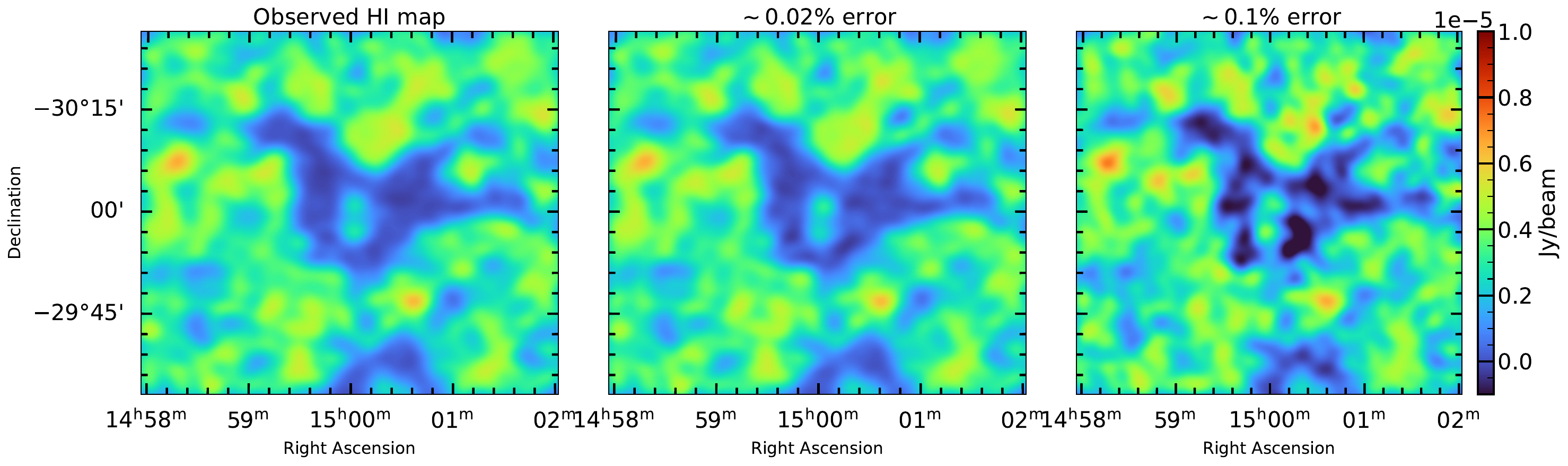}
    \includegraphics[width=\linewidth]{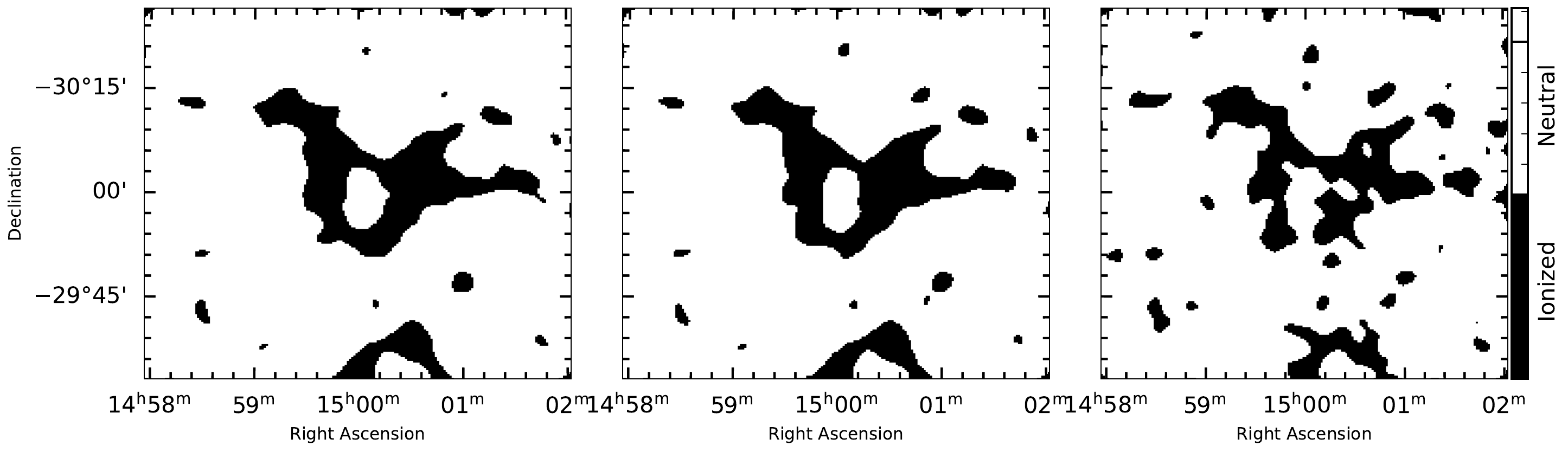}
    \caption{The visual representation of one such slice of the image cube at $\Bar{x}_{\text{HI}}\approx 0.55$. The observed \HI\ $21$\,cm brightness temperature map:  without any corruption (\texttt{Top Left}), with $\sim 0.02\%$ residual calibration errors (\texttt{Top Middle}), and  $\sim 0.1\%$ residual calibration errors (\texttt{Top Right}).  \texttt{Bottom:} The recovered \HII\ regions after applying the optimum thresholding method. The white (black) regions in these maps represent the recovered neutral (ionized) regions.}  
    \label{fig:gain-err_foregrounds}
\end{figure}

The calibration errors were applied by multiplying them with the model visibility data. The residual visibilities were then obtained by subtracting the foreground models using \textsc{CASA} {\sc uvsub} task. This {\sc uvsub}ed residual visibilities dataset is used to make image cubes for LCS analysis. The top panel of Figure \ref{fig:gain-err_foregrounds} shows the slices of the observed brightness temperature map at $\Bar{x}_{\text{HI}}\approx 0.55 (z = 7.76$). The top-left panel shows the observed \HI\ map without any bias due to corruption. The middle and right top panel shows the residual maps after introducing calibration errors of $\sim 0.02\%$ \& $\sim 0.1\%$, respectively. The bottom panel shows the recovered \HII\ regions identified using the optimum thresholding method. The white (black) regions in these maps represent the recovered neutral (ionized) regions. At the higher calibration errors level (calibration inaccuracy of $\sim 0.1\%$), residual foreground contamination introduces artificial filamentary or tunnel-like features (deconvolution artefacts) into the obtained 21\,cm observation images.

\subsection{Instrumental noise}
\label{sec:telescope_noise}
In this Section, we discuss how the sensitivity of the system, in conjunction with total integration time, affects the synthesized map. We added an uncorrelated Gaussian noise to the simulated visibilities. This is achieved by adding randomly generated values selected from a zero-mean Gaussian distribution to the complex visibility amplitudes for each frequency channel, time integration, baseline, and polarization. The amplitude of the thermal noise per baselines following the radiometer equation \citep{Taylor_1999} is given by
\begin{equation}
    \sigma_{\rm N} = \frac{2 k_B T_{\rm sys}}{ A_{\rm eff}\sqrt{\delta\nu \delta t}} [\rm Jy]
\end{equation}
Where $T_{\rm sys}$ is the system temperature, $A_{\rm eff}$ is the effective area of the antenna/station, $k_B$ is the Boltzmann constant, $\delta\nu$ is the frequency resolution and $\delta t$ is the integration time of the visibilities. The $ A_{\rm eff}/ T_{\rm sys}$ \footnote{\url{https://www.skao.int/sites/default/files/documents/SKAO-TEL-0000818-V2_SKA1_Science_Performance.pdf}} values for SKA1-Low are listed by Braun et al. \citep{Braun2019}. We interpolated the $ A_{\rm eff}/ T_{\rm sys}$ values to the frequencies of interest. Our previous study by Dasgupta et al. \citep{Dasgupta2023} introduced additive Gaussian noise in the image plane. The value of systematic noise is drawn from a zero-mean Gaussian random distribution with a specified standard deviation and added to the pixel values of an image. In contrast, this work uses a more realistic representation of instrumental noise in radio interferometric observations by introducing an uncorrelated Gaussian noise in the visibility domain. Furthermore, while Dasgupta et al. \citep{Dasgupta2023} applied a constant RMS noise value to the image, our method accounts for the system noise RMS as a function of observing frequency, reflecting the actual variation of noise characteristics across the observational bandwidth. For this case, we assume the residual foreground contaminations are below the \HI\ signal level. The only contamination present in the \HI\ maps is thermal noise. The contribution of the thermal noise in the visibility domain is rescaled by the factor of $\sqrt{t_{\text{obs}}/t^{\text{uv}}_{\text{obs}}}$, where $t^{\text{uv}}_{\text{obs}} = 2$\,h represents the observation time per day will be restricted to $\pm 1$\,hours around the transit time of the target EoR field and $t_{\text{obs}}$ is the total integration time. This rescaling represents the coherent averaging of visibility data over the total integration time. For this simulation, we vary the total integration time of $1000$, $1500$, and $2000$\footnote{Our simulations track the sky for only 2\,hours per observation for computational efficiency. To accumulate $2000$\,hours of observation time, this requires $1000$\,repetitions. In contrast, actual observations with a 4\,hour daily tracking time would necessitate $500$\,repetitions of the same sky patch, taking approximately two years to achieve the desired observation time. If we shift the phase centre a bit of arcsec, we can mitigate the systematics effect.} to achieve a good signal-to-noise ratio level (SNR $\gtrapprox 3$) in the synthesized map. The SNR of the image is defined as the ratio of the rms fluctuation of the target signal in the image cube to the rms thermal noise limit for SKA1-Low, evaluated over the total observation time and on the scale of the angular resolution element ($\theta$).

\begin{figure}[ht!]
    \centering
    \includegraphics[width=\linewidth]{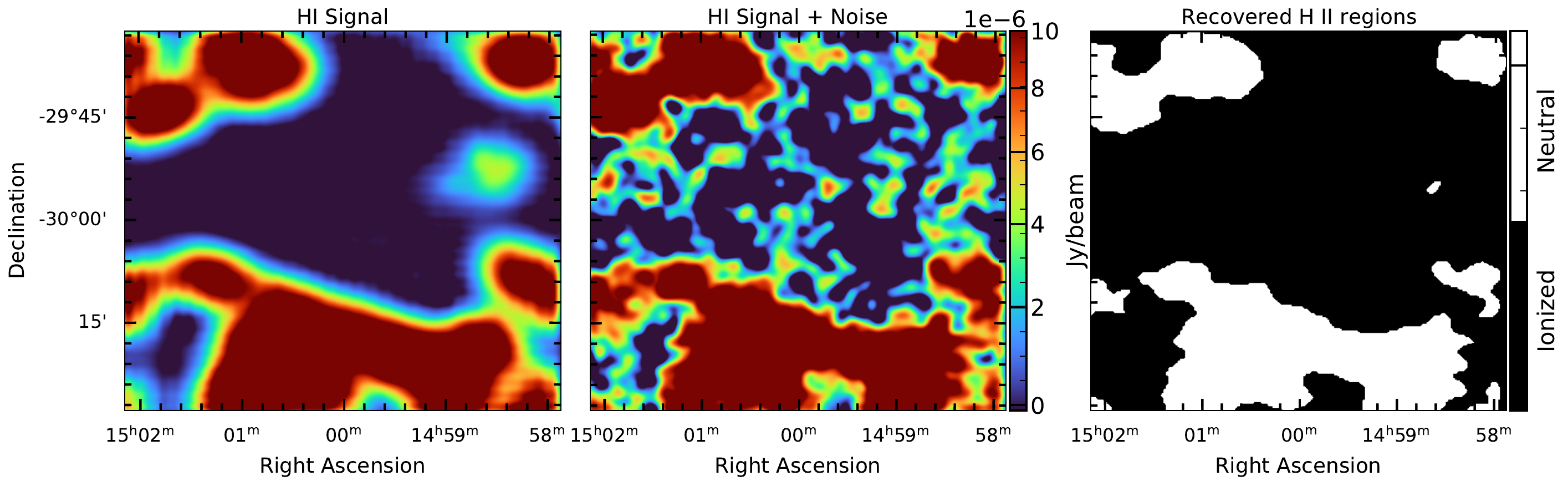}
    \caption{The pictorial representation of \HI\ maps at $\Bar{x}_{\rm HI}=0.2$ after performing multiscale cleaning with the natural weighting scheme through the \textsc{21cmE2E}-pipeline. \texttt{Left:} The observed \HI\ field without adding noise. \texttt{Middle:} Instrumental noise added to the \HI\ $21$\,cm field for an observation time of $2000$\,hours when observed with SKA1-Low with $296$ stations. \texttt{Right:} Identified ionized region from the $21$\,cm field after applying optimum thresholding algorithm. The white (black) regions in these maps represent the recovered neutral (ionized) regions.}
    \label{fig:thermal_noise}
\end{figure}

To incorporate the thermal noise into the simulation, we downsampled the simulated 21\,cm maps by a factor of $16$ along each side. As discussed in our previous study, the synthetic $21$\,cm observations require input maps with dimensions of $2^{3n}$ grids. We used a downsampling algorithm available in the \textsc{python} package \textsc{scikit-image} and integrated this method into the \textsc{21cmE2E} pipeline. This downsampling is essential to reduce the total observation time to achieve the desired SNR level. After downsampling the maps, the grid size of the final maps is $8.96$\,cMpc for redshift $z = 7.221$, which corresponds to an angular resolution of $3.45$\,arcmin and a frequency resolution of $0.54$\,MHz. We generated the image cubes using the \textsc{21cmE2E} pipeline after adding the thermal noise. In order to balance the resolution and the sensitivity of the telescope, two different imaging weighting schemes are used: natural weighting and Briggs weighting. The natural weighting scheme provides a relatively higher SNR but a lower resolution compared to the briggs weighting scheme with a robust parameter of $0.4$. The left and middle panels of Figure \ref{fig:thermal_noise} shows the slices of the brightness temperature field before and after adding the thermal noise with an observation time of $2000$\,h at a channel width of $0.54$\,MHz, respectively, at neutral fraction $\Bar{x}_{\text{HI}} = 0.2$. It is seen that more tunnel-like artefacts are visible on the map after introducing thermal noise. These artefacts result from the effect of noise and deconvolution. In the later section \ref{results_noise}, we discuss how this feature will affect the percolation process. The right panel of Figure \ref{fig:thermal_noise} shows the recovered \HII\ regions after applying an optimum thresholding method. In this map, black(white) regions represent recovered ionized (neutral) regions. 

\section{Percolation Transition \& LCS}
\label{sec:method}
In this section, we discuss a probe for the percolation process called the Largest Cluster Statistics (LCS) \cite{Klypin1993, Yess1996, Shandarin:1997fc}. In the early stage of reionization, a small group of ionized regions is formed, and the size and number of the ionized region grow gradually with time. At some point in time, these isolated ionized regions abruptly merge together to form a singly connected large ionized region spanning the entire IGM. This phase transition of the ionized region is known as the percolation transition \cite{zeldovich1982, Shandarin1983}. With the help of LCS, we draw inferences on the percolation process, especially on the percolation transition point. We identify the percolation transition occurring when the largest ionized region\,(LIR) spans the entire simulated volume and becomes formally infinitely extended owing to the periodic boundary conditions. In this work, which is a follow-up of a \citep{Bag_2018, Bag_2019, Pathak_2022, Dasgupta2023}, we follow the evolution of the large ionized region as reionization advances using LCS. The LCS is defined as 
\begin{equation}
    \text{LCS}  = \frac{\text{volume of the largest ionized region}}{\text{total volume of all the ionized regions}}
    \label{lcs_defination}
\end{equation}
As can be seen from the above definition, LCS represents the fraction of ionized volume residing within the LIR. Hence, at the onset of the percolation transition, an abrupt increase in LCS is anticipated. This abrupt transition defines the percolation transition threshold. We plot LCS as a function of the mass-averaged neutral fraction ($\Bar{x}_{\text{HI}}$), e.g. in figure \ref{fig:lcs_signal}, to characterize the evolution of the LIR as the neutral fraction decays with reionization. The critical $\Bar{x}_{\text{HI}}$ denotes the threshold at which percolation transition occurs, where small ionized regions merge into a large ionized region spanning the entire IGM. The sudden increase in LIR volume results in a sharp increase in the LCS. We identify the percolation transition threshold as the point at which the change in LCS is the maximum. Therefore, both the LCS profile and its critical value at the percolation transition serve as crucial metrics for evaluating the morphological evolution and history of reionization. Previous works by \citep{Bag_2018, Bag_2019, Pathak_2022, Dasgupta2023} have shown that the LCS can be a good metric to probe the percolation process and thereby distinguish extreme models of reionization. Our study focuses on the evolution of LCS from the coeval boxes observed through synthetic 21\,cm observations with SKA1-Low. In real observation, we observed the light-cone effect \cite{Datta2012}. However, its impact on the evolution of the LCS along a line-of-sight has minimal (Potluri et al., in prep). We use coeval boxes to study the evolution of LCS from 21\,cm observation. Therefore, the effect of light-cone is not expected to be so dominant and is beyond the scope of this paper.

\subsection{Binarization of the image cubes}
\label{image_cube}
In order to estimate the LCS on the brightness temperature maps, we use a code, \textsc{SURFGEN2} \citep{Sheth:2002rf, Sheth:2006qz, Bag_2018, Bag_2019}. \textsc{SURFGEN2} not only determines the LCS but also helps identify the topological and morphological features of each ionized region using Shapefinders \cite{Sahni:1998cr}. \footnote{Shapefinders are derived from the ratios of the Minkowski functionals. For instance, in three dimensions, the four Minkowski functionals give rise to three shapefinders, with each one representing the extent of a closed surface along one of the three dimensions; see \cite{Sahni:1998cr} for more details.} For an in-depth understanding of the operating principles of the \textsc{SURFGEN2} code, readers can refer to \citep{Sheth:2006qz, Bag_2019}. The \textsc{SURFGEN2} code requires thresholding to binarize the neutral and ionized regions of the \HI\ maps. In an ideal \HI\ 21\,cm brightness temperature field ($\delta T_b$), an ionized region is identified by where $\delta T_b$ equal to zero. For an interferometric observation, we measure only the fluctuations in the $21$\,cm signal. Accordingly, the minima of the brightness temperature maps correspond to ionized regions, as expected. However, in real observations, the presence of systematic effects poses a challenge in finding the optimal threshold to distinguish the ionized regions. In our previous work, Dasgupta et al. \cite{Dasgupta2023}, imposed a gradient-descent method on the histograms of the synthetically observed \HI\ maps from the SKA1-Low to binarize them. When we add calibration errors and instrumental noise to the \HI\ maps, finding an optimal threshold becomes a challenging task. 

In this work, we developed a modified approach to identify ionized regions in the 21\,cm observation maps by combining a global thresholding method with the unsharp masking technique. Unsharp masking is a widely used image enhancement method in image processing that sharpens features by emphasizing edges and fine details. It is a linear operation, making it a preferred choice over deconvolution, which is often an ill-posed problem. This technique enhances edges (and other small-scale features in an image) by subtracting an unsharp or smoothed (blurred) version of the image from the original, thereby highlighting small-scale structures. The unsharp masking operation is mathematically expressed as:
\begin{equation}
    f_{\rm sharp}(x,y) = f(x,y) + k ~[f(x,y) - f_{\rm smooth}(x,y)] 
\end{equation}
where $f(x,y)$ is the original image, $f_{\rm smooth}(x,y)$ is its smoothed version, and $k$ is a scaling factor that controls the strength of the enhancement. A higher value of $k$ increases the contrast of fine details, while a lower value results in more subtle sharpening. We applied the unsharp masking algorithm using the implementation from the \textsc{scikit-image} Python package. This step better defines the boundaries of ionized bubbles prior to applying a global thresholding criterion for more precise segmentation. We apply the method of optimum thresholding to binarize the observed 21\,cm maps produced by the \textsc{21cmE2E}-pipeline and calculate LCS for each of the image cubes.

To assess the effectiveness of our method, we compare it with the binarization technique described in Giri et al. \citep{Giri_2025arXiv}. We observed that their approach tends to over-segment ionized regions in the presence of low-density fluctuations within neutral regions. However, our method demonstrates robustness across diverse density environments and more accurately preserves the morphological structure of ionized regions (see Appendix~\ref{binarization_image}). It is important to note that the binarization technique can be biased by the instrumental noise. The instrumental noise introduces the small-scale features in the HI maps from 21\,cm observation. In order to mitigate this, a Gaussian filter is applied prior to our modified binarization scheme. This step ensures robust segmentation from noisy 21\,cm observation data. This method, herein termed optimal thresholding, is proposed as the robust thresholding method in all scenarios. In future work, we will also investigate the optimal thresholding for binarizing the \HI\ field from interferometric observations (Dasgupta et. al., in prep).


\section{Results}\label{sec:results}
This section discusses the impact of the thresholding method, antenna-based calibration errors and the instrumental noise for SKA1-Low on LCS analyses. We use the \textsc{21cmE2E} pipeline to assess how each of these factors affects the percolation process during different stages of reionization. The following subsections present detailed results for each factor considered.

\subsection{Effect of thresholding}\label{thresholding}
This work presents an optimum strategy to reduce the bias in LCS estimation from the simulated 21\,cm observation results. Our previous study done by Dasgupta et al. \citep{Dasgupta2023}, used a thresholding method, e.g., gradient descent, to binarize 21\,cm maps and estimate the LCS. During the early stages of reionization, the histogram of 21\,cm maps is not bimodal. Instead, it exhibits an unimodal distribution with asymmetric tails, as indicated by non-zero skewness \citep{Ilieve_2006, Watkinson_2015}. This thresholding method introduces bias for these cases due to its limitations in reaching the local minima in the 21\,cm observation maps. The problem becomes more significant when telescope effects, such as the proper implementation of the array synthesized beam of SKA1-Low, are included. Therefore, these limitations fail to binarize the image pixels and lead to biases in the recovered percolation process of reionization.

\begin{figure}[ht!]
    \centering
    \includegraphics[width=\linewidth]{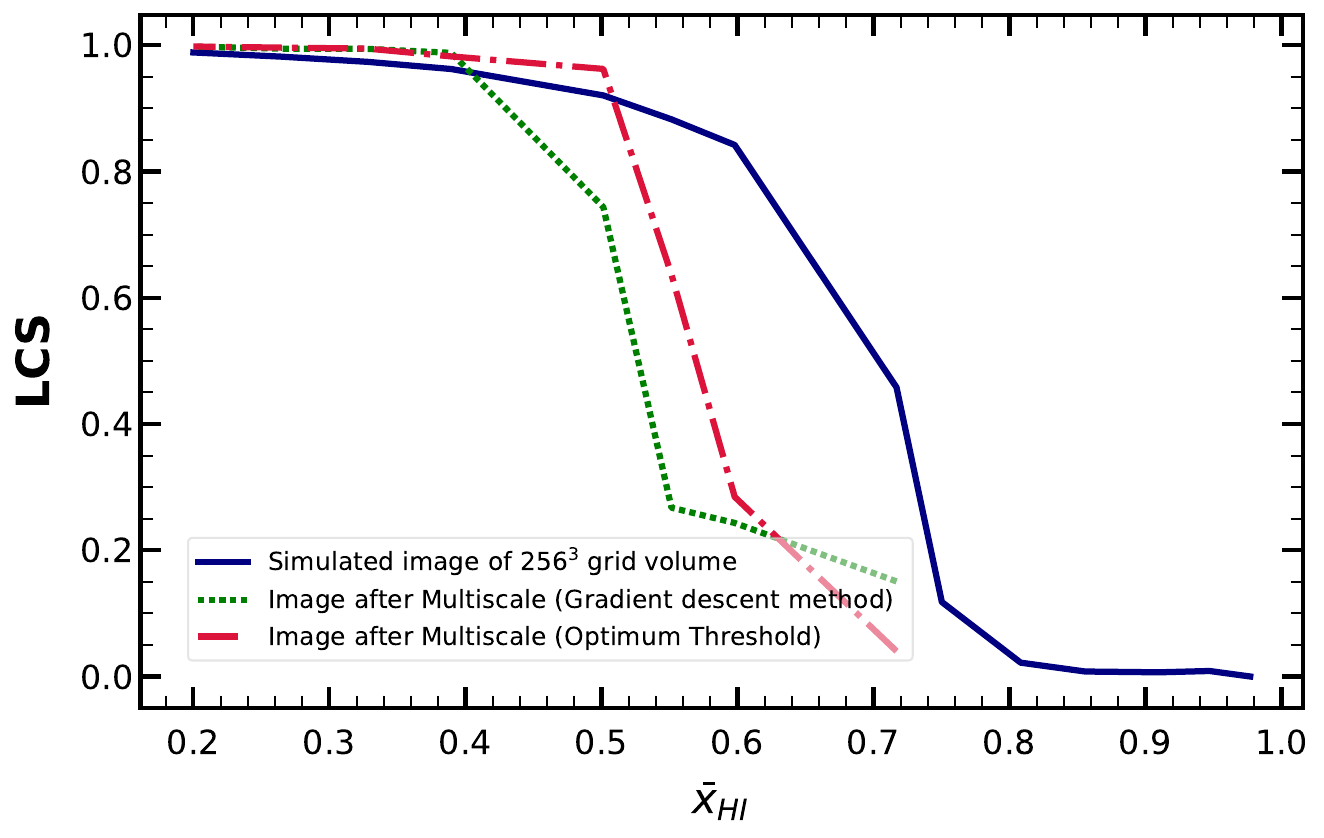}
    \caption{Comparison of the obtained LCS from 21\,cm observation maps against neutral fraction for different thresholding methods. The original image is a hypothetical scenario without telescope effect and noise, and the corresponding threshold for LCS is set at zero. The red dash-dot and green dashed curves illustrate the obtained LCS based on the threshold set by the optimum thresholding and gradient-descent methods. The optimum thresholding method demonstrates superior performance compared to the gradient-descent method.}
    \label{fig:lcs_signal}
\end{figure}

In order to mitigate these issues, we propose an optimum thresholding method for robustly separating neutral and ionized regions. In figure  \ref{fig:lcs_signal}, we plot the comparison of the obtained LCS from the simulated 21\,cm observation against neutral fraction using different thresholding methods. The original image is a hypothetical scenario without telescope effect and noise, and the corresponding threshold for LCS is set at zero. The red dash-dot and green dashed curves represent the estimation of LCS based on the thresholds obtained via the optimum thresholding method and gradient-descent methods, respectively. The optimum thresholding method demonstrates superior performance compared to the gradient-descent method. We observe that the threshold set by the optimum thresholding method on the simulated 21\,cm observations results in the same percolation transition redshift or $\Bar{x}_{\text{HI}} \approx 0.7$ \citep{Ilieve_2006, Furlanetto_2016, Bag_2018, Pathak_2022, Dasgupta2023} as that of the hypothetical scenario, i.e. without any telescope effect and noise shown by the blue solid curve. However, the obtained LCS based on the threshold set by the optimum thresholding method has deviations from the original LCS estimation, due to the resolution of the telescope and the error in the thresholding method.

\begin{figure*}[hb!]
    \centering
    \includegraphics[width=\linewidth]{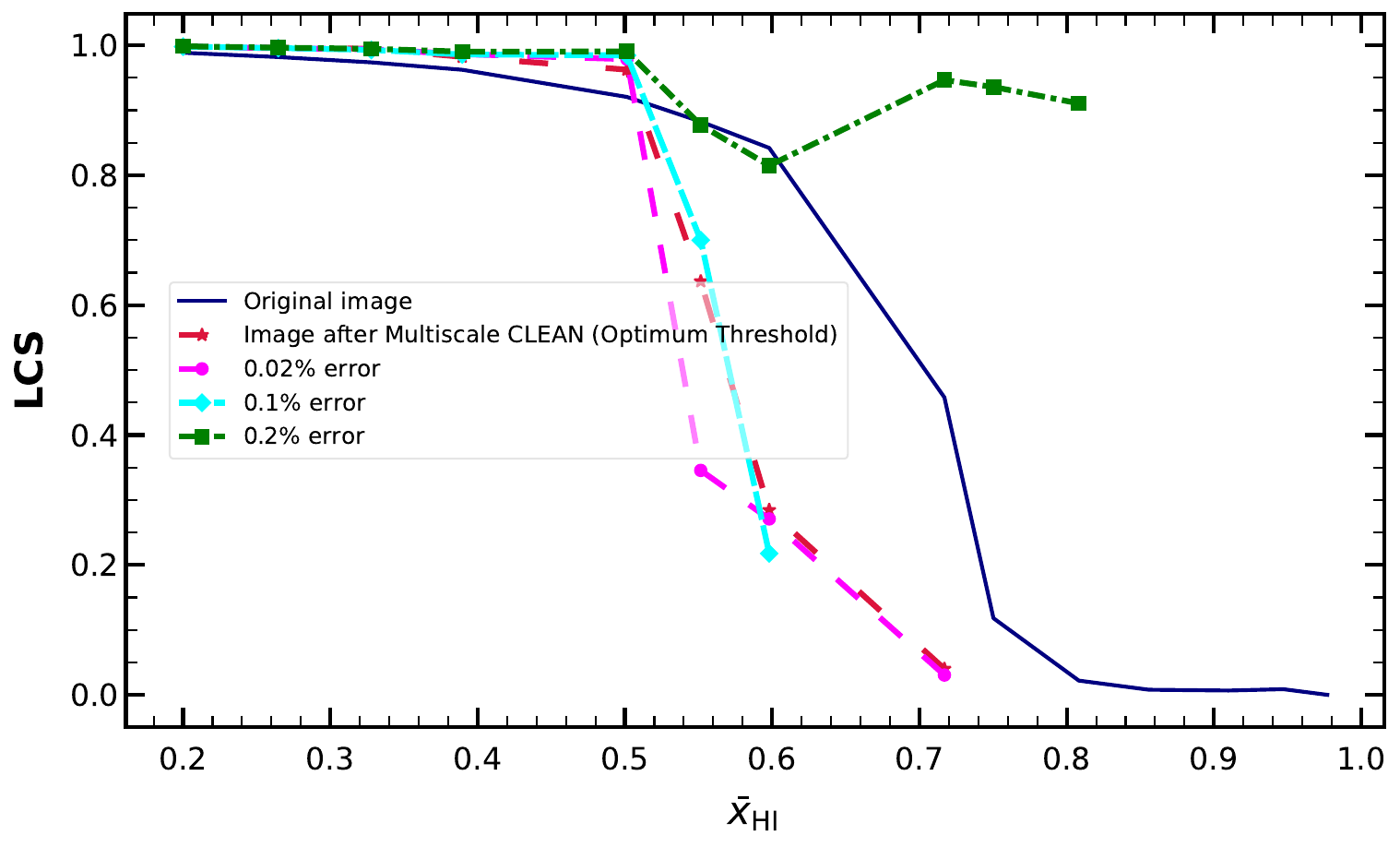}
    \caption{Comparison of the obtained LCS as a function of $\Bar{x}_{\text{HI}}$, for different gain calibration errors on simulated SKA1-Low observational maps (Case I). These maps include both unsubtracted point source contamination and the target 21\,cm signal. It is observed that with residual post-calibration inaccuracy of $0.02\%$ (magenta circle), the obtained LCS remains unaffected mainly and closely matches the evolution of the LCS in the 21\,cm observations without corruption.}
    \label{fig:lcs_point}
\end{figure*}

\subsection{Effect of gain calibration error}
\label{results_gain}
In this section, we discuss the impact of time-frequency correlated antenna-based residual gain calibration errors on the obtained LCS from 21\,cm observation results. This LCS analysis is based on the threshold set via the optimum thresholding method on residual image cubes. There are many techniques that have been developed in the past decade to remove residual contamination from the image domain \citep{Mertens_2020}. However, we aim to determine the maximum tolerance level of antenna-based calibration error to recover the reionization history in a robust and relatively unbiased manner using LCS. In order to estimate the tolerance level, we increase the level of calibration errors. This approach helps us ascertain the dynamic range required to extract the 21\,cm signal effectively from the residual data. For this work, we consider two types of foreground contamination in our LCS analysis, as detailed in Section \ref{foregrounds}. The results obtained for these different foreground models are discussed in the subsequent subsections.
    
\textbf{Case I}: Figure~\ref{fig:lcs_point} compares the evolution of the obtained LCS with $\Bar{x}_{\text{HI}}$ for varying levels of residual calibration errors on the simulated SKA1-Low observational maps. These maps include both unsubtracted point source contamination and the target 21-cm signal. The effect of residual calibration errors on LCS estimation is shown in magenta circle, cyan diamond, and green square curves in figure~\ref{fig:lcs_point}. These results are obtained from the simulated 21\,cm observation via \textsc{21cmE2E} pipeline based on the threshold set by the optimum thresholding method. The blue solid curve represents the original image, a hypothetical scenario without telescope effect and noise, and the corresponding threshold for LCS is set at zero. We observed that with a residual gain calibration inaccuracy of $0.02\%$ (illustrated by the red star curve), the obtained LCS from simulated 21\,cm observations remains largely unaffected. It closely matches the evolution of the LCS  as seen in the obtained LCS from 21\,cm observations without any bias due to corruption. 

However, with higher calibration errors, the computation of LCS from the simulated SKA1-Low observational maps becomes challenging. This occurs because the residual foreground, resulting from residual gain calibration errors, introduces artificial filamentary or tunnel-like features (deconvolution artefacts) in the final image cube. These artefacts lead to the fragmentation of the largest ionized region into isolated regions. The extent of this fragmentation depends on the thresholding method used to binarize each \HI\ map. Furthermore, this systematic effect reduces the contrast between ionized and neutral pixels. This reduction is primarily due to the additional RMS noise introduced by the residual contamination, which in turn lowers the overall signal-to-noise ratio (SNR). Therefore, this contamination makes it challenging to identify the ionized and neutral regions accurately and leads to a biased interpretation of the history of reionization using LCS. 

\begin{figure}[hb!]
    \centering
    \includegraphics[width=\linewidth]{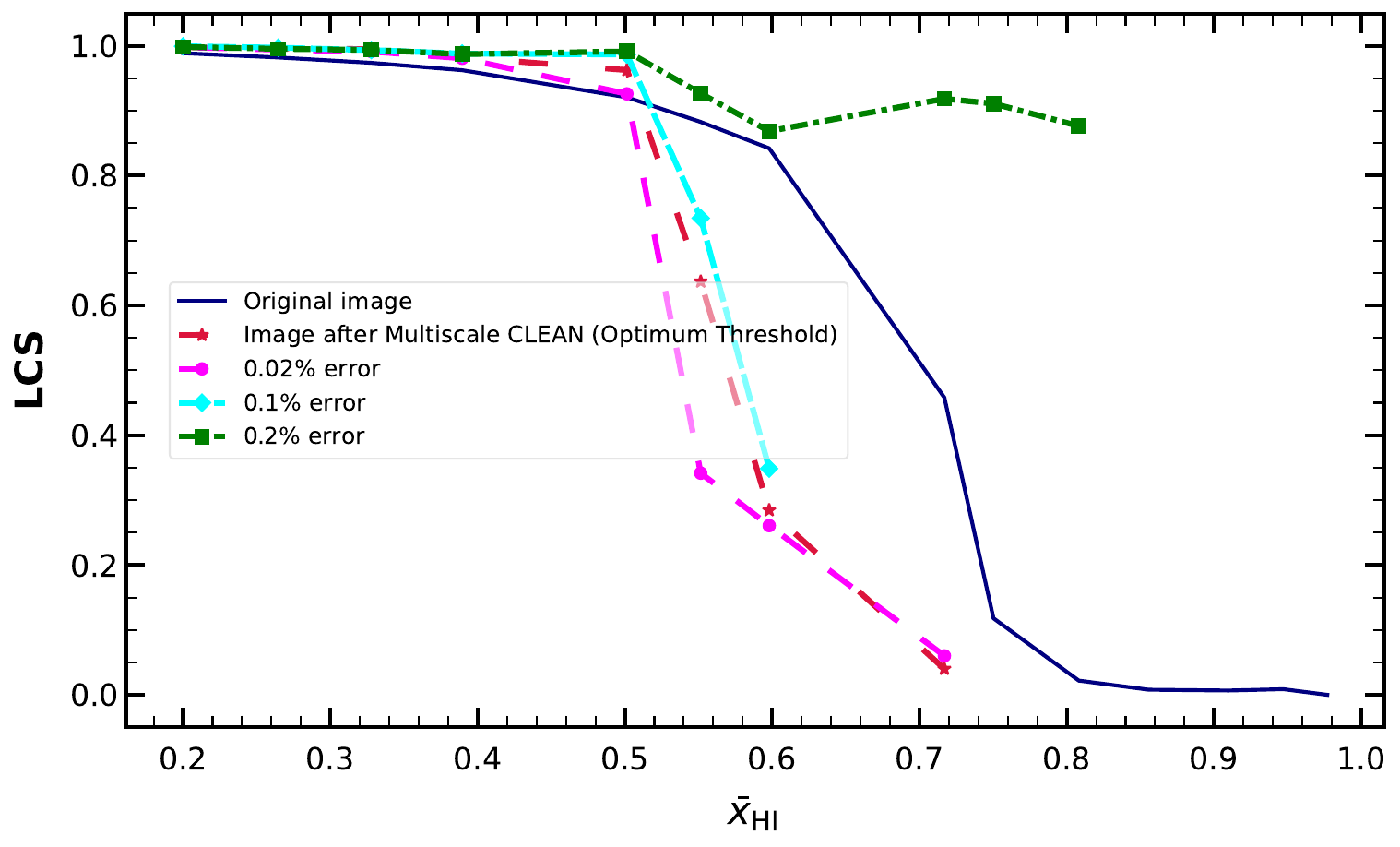}
    \caption{Same as figure~\ref{fig:lcs_point} for Case II. The simulated SKA1-Low observational maps include unsubtracted point source contamination, diffuse emission and the target signal.}   
    \label{fig:lcs_foregrounds}
\end{figure}

\textbf{Case II:} Figure~\ref{fig:lcs_foregrounds} compares the evolution of the obtained LCS with $\Bar{x}_{\text{HI}}$ for different levels of residual calibration errors on the simulated SKA1-Low observational maps for Case II. These maps include unsubtracted point source contamination, diffuse emission and the target signal. The magenta circle, cyan diamond, and green square curves illustrate the effect of the residual post-calibration errors on LCS estimation in figure~\ref{fig:lcs_foregrounds}. Similar to Case I, we also observed with the residual gain calibration, the inaccuracy of $0.02\%$ (illustrated by the red star curve), the obtained LCS from simulated 21\,cm observations remains largely unaffected. It closely matches the evolution of the LCS  as seen in the obtained LCS from 21\,cm observations without any bias due to corruption. Therefore, a post-calibration inaccuracy of $0.02\%$ is required to recover the history of reionization in a robust and relatively unbiased manner using LCS. Although a calibration error of $0.1\%$, the apparent percolation curves visually appear to match with a zero-error case. However, this small error can still substantially contribute to image artefacts and morphological distortions. 

It is essential to note that imperfect subtraction of bright sources can also create negative bowl-like regions. These regions further fragment the largest ionized regions into isolated pieces, thereby introducing additional bias into the LCS analysis. Our initial systematic study emphasizes the critical importance of achieving high calibration accuracy for the upcoming SKA1-Low observations. Such accuracy is essential to recover the reionization history in a robust and unbiased manner using the LCS. However, the choice of thresholding algorithm significantly impacts the LCS analysis. Future work will explore various thresholding algorithms to improve the separation of neutral and ionized regions during LCS estimation.

\begin{figure}
    \centering
    \includegraphics[width=\linewidth]{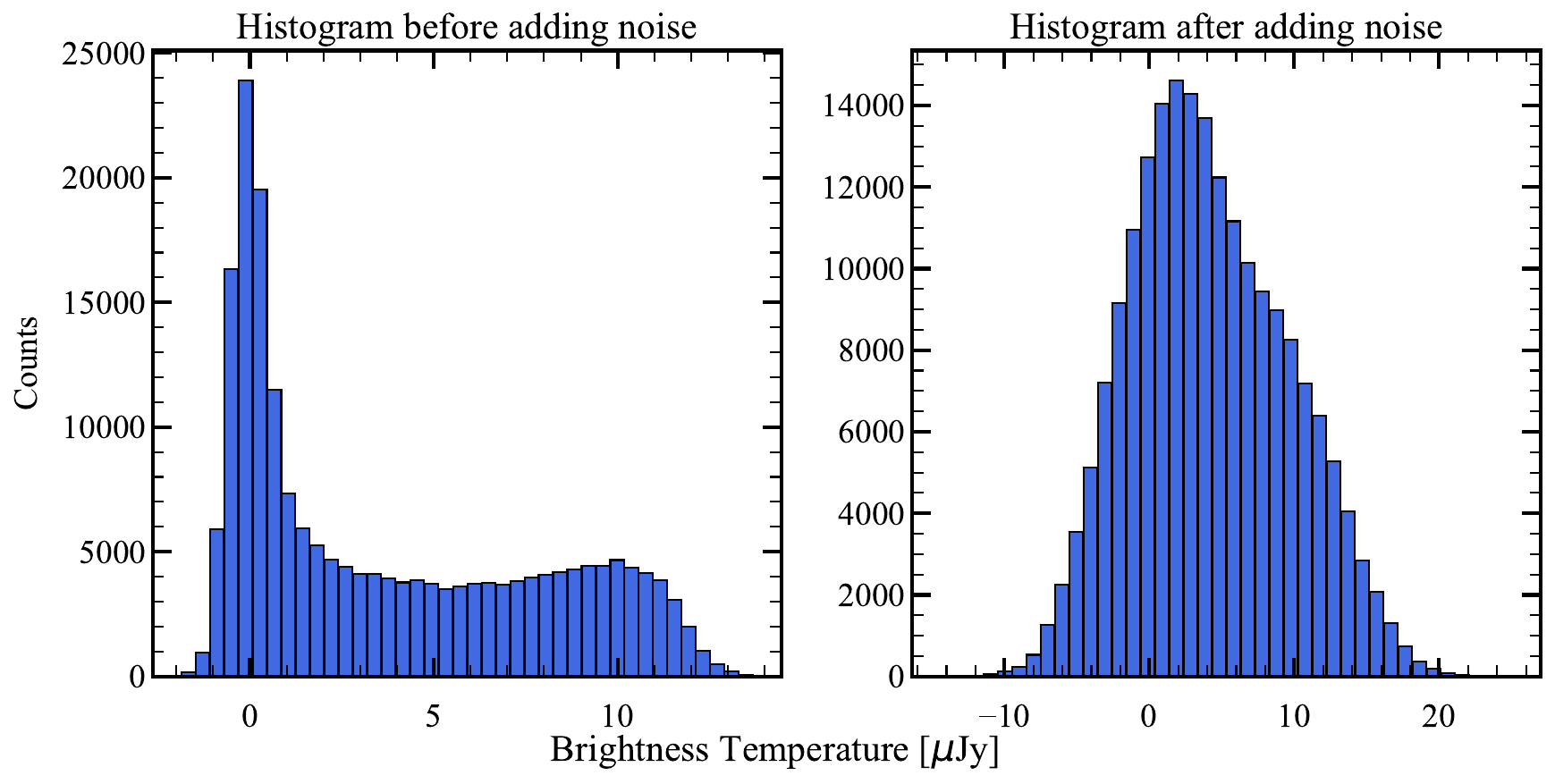}
    \caption{The variation in the histogram of bimodal distributions of 21\,cm field at a fixed neutral fraction before (left panel) and after (right panel) the addition of SKA1-Low instrumental noise for 2000\,hours of deep integration time. Introducing random fluctuations into the simulated 21\,cm signal causes the loss of bimodality, shifting the histogram toward a Gaussian distribution. This obscures the sharp boundary between ionized and neutral regions, making it challenging to accurately binarize the image pixels.}
    \label{fig:hist_noise}
\end{figure}

\subsection{Effect of instrumental noise}\label{results_noise}
\begin{figure}
    \centering
    \includegraphics[width=\linewidth]{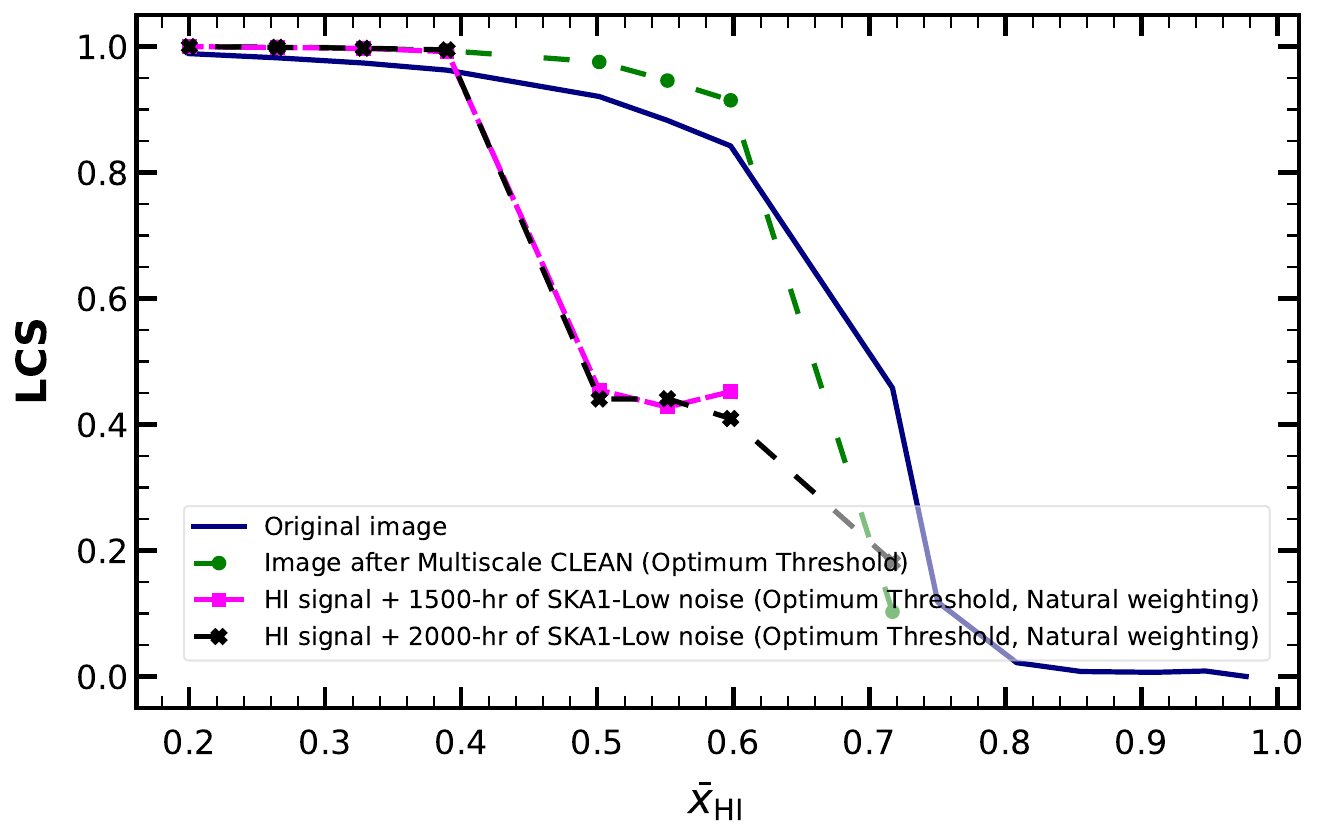}
    \caption{LCS is computed as a function of neutral fraction $\Bar{x}_{\text{HI}}$ for varying levels of accumulated time of observations. It is observed that the apparent percolation process systematically shifts towards a lower $\Bar{x}_{\text{HI}}$. This shift is caused by the combined effects of random noise fluctuations and the introduction of partially ionized regions within the HI 21\,cm field. This can lead to a significantly biased interpretation of the reionization history.}
    \label{fig:lcs_noise_yen}
\end{figure}

In this section, we discuss the effect of the instrumental noise using a realistic simulation on the synthesized SKA1-Low observational maps. In section ~\ref{sec:telescope_noise}, we discuss the prescription to add instrument noise in the visibility domain. Figure~\ref{fig:lcs_noise_yen} illustrates the evolution of the obtained LCS from 21\,cm observation against $\Bar{x}_{\text{HI}}$ for different deep observation times. The original image is a hypothetical scenario without telescope effect and noise, and the corresponding threshold for LCS is set at zero, indicated by a blue solid curve. A green dotted curve illustrates the obtained LCS from the observed downsample image, based on the threshold set by the optimum thresholding method. For deep 21\,cm observations with 1500\,hours and $2000$\,hours of integration time, the obtained LCS (threshold set by the optimum thresholding method) is shown by the magenta and black curves, respectively. It is observed that, although the LCS features can be computed up to a certain  $\Bar{x}_{\text{HI}}$,  the percolation transition threshold systematically shifts towards a lower $\Bar{x}_{\text{HI}}$. This systematic shift arises from the combined effects of random noise fluctuations and the introduction of partially ionized regions within the HI 21\,cm field. Consequently, such a systematic bias in the percolation transition threshold may lead to a significantly biased interpretation of the reionization history. This similar feature of LCS is observed when considering the combined effects of additive Gaussian noise in the image plane and the choice of larger smoothing scales \citep{Dasgupta2023}.

Our findings highlight the significance of imaging weighting to obtain the LCS from 21\,cm observation. In addition, it emphasized the importance of the thresholding algorithm to separate neutral and ionized pixels in noisy data and the impact of cluster detection within the \textsc{SURFGEN2} code on LCS estimation. Future work will explore alternative thresholding algorithms for improved pixel differentiation.

\section{Summary and discussion}
\label{sec:summary}
This work investigated the effect of antenna-based calibration errors and instrumental noise for SKA1-Low on LCS analysis. We aim to recover the reionization history using the LCS in a robust and relatively unbiased manner. The key findings of this investigation are summarized as follows: 
\begin{itemize}
    \item The histogram of the brightness temperature map of the 21\,cm field during EoR exhibits a bimodal distribution. The two peaks correspond to ionized and neutral regions. In a hypothetical scenario without telescope effects or noise, ionized regions would have a brightness temperature of zero. However, in radio interferometric observation, determining an optimal threshold to distinguish neutral from ionized regions in the 21\,cm brightness map is challenging. The complexity arises from the systematic effects, resolution of the telescope, and thermal noise, which cause random shifts in pixel brightness temperature maps. Therefore, accurately differentiating between these regions in 21\,cm maps becomes difficult. In this study, we introduce an optimal thresholding strategy to binarize the image pixels and accurately recover the percolation transition. It is observed that this method significantly reduces the bias in LCS estimation from 21cm maps. However, the obtained LCS has little deviation from the actual LCS estimation from the hypothetical scenario due to the resolution of the SKA1-Low telescope and error in the thresholding method.
    
    \item We study the effect of direction-independent calibration errors on the percolation process of reionization history using the LCS of the 21\,cm maps. We have demonstrated that a post-calibration and post-averaging antenna-based calibration error tolerance of $\sim 0.02\%$ is essential to achieve unbiased and unaffected LCS estimation. This corresponds to an amplitude error of $\sim 0.02\%$  and a phase error of $\sim 0.02^{\circ}$ in each time domain. It is important to note that the tolerance of $\sim 0.02\%$ will vary depending on the RMS variation across different sky patches, i.e., the statistical distribution of bright sources within the FoV. We also observed that the threshold set by the optimum thresholding method on the residual 21\,cm maps simulated via \textsc{21cmE2E} aligns with the same percolation transition redshift or $\Bar{x}_{\rm{HI}} = \sim 0.7$, as results from the simulated 21\,cm observations without any bias due to corruption. However, at higher calibration errors( $\sim 0.2\%$), the dynamic range of the 21\,cm maps significantly deteriorates. The RMS noise from the residual foreground starts to dominate the image pixels and introduces spurious features that can either mimic or entirely obscure the faint 21\,cm cosmological signal. Under these conditions, the thresholding method,  fails to identify the boundary between the ionized and neutral pixels. This results in a biased estimation of LCS at the higher calibration errors.

    \item Next, we introduce the instrumental noise for SKA1-Low to assess the robustness of the evolution of the obtained LCS. After introducing the instrumental noise, it is observed that the bimodal feature of the 21\,cm map is lost, and the histogram shifts towards a Gaussian distribution. We observed that with an accumulation of $2000$\,h of observation time, the estimation of the LCS threshold set by the optimum thresholding method, systematically shifts towards a lower $\Bar{x}_{\text{HI}}$. This systematic shift arises from the combined effects of random noise fluctuations and the introduction of partially ionized regions within the HI 21\,cm field. In the early stages of the EoR (i.e., at higher neutral fractions), the limited resolution causes small ionized regions to be confused with partially neutral within the \HI\ 21\,cm field. Therefore, at this stage, the estimation of LCS could not be computed. As the contrast between ionized bubbles and random fluctuations becomes comparable, the performance of the LCS analysis is affected. At the later stage of reionization, the combination of poor resolution and instrumental noise leads to a biased determination of the threshold between ionized and neutral regions. This, in turn, results in erratic behaviour in the evolution of the recovered LCS from 21\,cm observations and biased interpretation of reionization history. This feature of LCS is observed when considering the combined effects of additive Gaussian noise in the image plane and the choice of larger smoothing scales \citep{Dasgupta2023}.
    
\end{itemize}

Our findings demonstrate that the obtained LCS results can be significantly biased by the choice of thresholding method. This highlights the critical need for developing a robust thresholding technique to ensure precise percolation analysis across all reionization stages. However, we acknowledge that our current study did not account for the impact of ionospheric phase distortion, the chromaticity of the primary beam, and other systematic instrumental effects. These factors further complicate LCS analysis and potentially introduce additional biases. Our future work will address these factors to provide a more comprehensive understanding of their implications for EoR $21$\,cm observations, and subsequent LCS analysis.

\section{Acknowledgements}
SKP acknowledges the financial support by the Department of Science and Technology, Government of India, through the INSPIRE Fellowship [IF200312]. The authors acknowledge the use of facilities procured through the funding via the Department of Science and Technology, Government of India sponsored DST-FIST grant no. SR/FST/PSII/2021/162 (C) awarded to the DAASE, IIT Indore. AD and SM acknowledge financial support through the project titled ``Observing the Cosmic Dawn in Multicolour using Next Generation Telescopes'' funded by the Science and Engineering Research Board (SERB), Department of Science and Technology, Government of India through the Core Research Grant No. CRG/2021/004025. SB acknowledges the funding provided by the Alexander von Humboldt Foundation. SB thanks Varun Sahni and Santanu Das for their contributions to the development of \textsc{SURFGEN2}.

\appendix
\section{Image Binarization}\label{binarization_image}
\begin{figure}
    \centering
    \includegraphics[width=\linewidth]{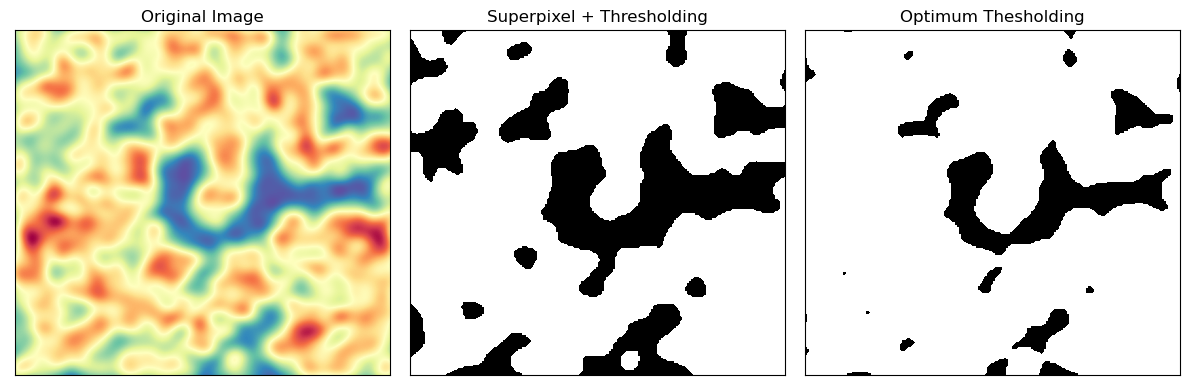}
    \caption{An illustration of the comparison of different thresholding methods for segmenting neutral and ionized pixels from simulated 21\,cm observation maps. Left: A simulated 21\,cm observation map. Middle: The binary image map produced by the binarization technique from Giri et al.\citep{Giri_2025arXiv}. They identified the structure by combining a superpixel and Li thresholding methods. Right: The binary image map resulting from the optimal thresholding method employed in this work.}
    \label{fig:thrs_comp}
\end{figure}

Threshold selection for identifying neutral and ionized pixels is a crucial aspect of EoR 21\,cm image analysis. This work introduces a modified method that combines a global thresholding strategy with an unsharp masking technique,  to identify ionized regions within 21cm observation maps. As illustrated in Figure~\ref{fig:thrs_comp}, we compare our method with the binarization technique described by Giri et al. \citep{Giri_2025arXiv}. We find that, whereas their method tends to over-segment ionized regions, particully in areas with low-density fluctuations. Our method demonstrates superior robustness across diverse density environments and more accurately preserves the morphological structure of ionized regions. The black (white) regions in binary maps represent the ionized and neutral regions.

\bibliographystyle{JHEP}
\bibliography{ref}

\end{document}